# Assessment of the Failure of Active Days Fraction Method of Sunspot Group Number Reconstructions

Leif Svalgaard[1], and Kenneth H. Schatten[2]

**Abstract**

We identify several pairs of 'equivalent' observers defined as observers with equal or nearly equal 'observational threshold' areas of sunspots on the solar disk as determined by the 'Active Days Fraction' method [e.g. Willamo et al., 2017]. For such pairs of observers, the ADF-method would be expected to map the actually observed sunspot group numbers for the individual observers to two reconstructed series that are very nearly equal and (it is claimed) represent 'real' solar activity without arbitrary choices and deleterious, error-accumulating 'daisy-chaining'. We show that this goal has not been achieved (for the critical period at the end of the 19[th] century and the beginning of the 20[th]), rendering the ADF-methodology suspect and not reliable nor useful for studying the long-term variation of solar activity.

## 1. Introduction

Willamo et al. [2017] upgrade the Sunspot Group Number reconstruction based on the fraction of 'Active Days' per month suggested by Vaquero et al. [2012] as extended by Usoskin et al. [2016] and touted as a "modern non-parametric method […] free from daisy-chaining and arbitrary choices". The method uses the ratio between the number of days per month when at least one group was observed and the total number of days with observations. This Active Days Fraction, ADF, is assumed to be a measure of the acuity of the observer and of the quality of the telescope and counting technique, and thus might be useful for calibrating the number of groups seen by the observer by comparing her ADF with a modern reference observer.

A problem with ADF is that near sunspot maximum, every day is an 'active day' so ADF at such times is nearly always unity and thus does not carry information about the statistics of high solar activity. This 'information shadow' occurs for even moderate group numbers greater than three. Information gleaned from low-activity times must then be extrapolated to cover solar maxima under the hard-to-verify assumption that such extrapolation is valid regardless of activity, secularly varying observing technique and counting rules, and instrumental technology.

In this article we test the validity of the assumptions using pairs of high-quality observers where within each pair the observers every year reported very nearly identical group counts distributed the same way for several decades. The expectation on which our assessment rests is that the ADF method shall duly reflect this similarity and yield very similar reconstructions, for both observers within each pair. If not, we shall posit that the ADF method has failed (at least for the observers under test) and that the method therefore cannot without qualification be relied upon for general use.

---

[1] Stanford University, Stanford, CA 94305, USA; email: leif@leif.org
[2] a.i. solutions, Lanham, MD 20706, USA

## 2. The Observers and Their Data

Winkler and Quimby form the first pair. Wilhelm Winkler (1842-1910) - a German private astronomer and maecenas [Weise et al., 1998] observed sunspots with a Steinheil refractor of 4-inch aperture at magnification 80 using a polarizing helioscope from 1878 until his death in 1910 and reported his observations to the Zürich observers Wolf and Wolfer who published them in full in the 'Mittheilungen' whence Hoyt & Schatten [1998] extracted the group counts for inclusion in their celebrated catalog of sunspot group observations[3]. The Reverend Alden Walker Quimby of Berwyn, Pennsylvania observed from 1892-1921 with a 4.5-inch aperture telescope with a superb Bardou lens (1889-1891 with a smaller 3-inch aperture). The observations were also published in full in 'Mittheilungen' and included in the Hoyt & Schatten catalog. As we shall see in Section 3, Winkler and Quimby have identical group $k'$-values with respect to Wolfer and thus saw and reported comparable number of sunspot groups.

Broger and Wolfer form a second pair. Max Broger (18XX-19ZZ) was hired as an assistant at the Zürich Observatory and observed 1896–1936 using the same (still existing) Fraunhofer-Merz 80mm 'Norm telescope' at magnification 64 as the director Wolfer. Alfred Wolfer (1854-1931) started as an assistant to Wolf in 1876 and observed until 1928. Broger had a $k'$-value of unity with respect to Wolfer and thus saw and reported comparable number of sunspot groups. In addition, there probably was institutional consensus as to what would constitute a sunspot group. The observations were direct at the eyepiece and all were published in the 'Mittheilungen' and from 1880 on in the Hoyt & Schatten catalog.

The original Hoyt & Schatten catalog has been amended and in places corrected and the updated and current version [Vaquero et al., 2016] is now curated by the World Data Center for the production, preservation and dissemination of the international sunspot number in Brussels: http://www.sidc.be/silso/groupnumberv3[4]. Ilya Usoskin has kindly communicated the data extracted from the above that were used for the calculation [Willamo et al., 2017] of the ADF-based reconstruction of the Group Number. We have used that selection (taking into account the correct Winkler 1892 data[3]) for our assessment (can be freely downloaded from http://www.leif.org/research/gn-data.htm).
We compute monthly averages from the daily data, and yearly averages from months with at least 10 days of observations during the year. It is very rare that this deviates above the noise level from the straight yearly average of all observations during that year.

---

[3] Unfortunately, the data in the original Hoyt & Schatten data files for Winkler in 1892 are not correct. The data for Winkler in the data file are really those for Konkoly at O-Gyalla for that year. L.S. has extracted the correct data from the original source [Wolf, 1893].
[4] Also available at http://haso.unex.es/?q=content/data

## 3. Winkler and Quimby are Equivalent Observers

Figure 1 shows that Winkler and Quimby have (within the errors) the same $k'$-factors (1.295±0.035 and 1.279±0.034) with respect to Wolfer, based on yearly values. For monthly values the factors are also equal (1.25±0.02 and 1.27±0.02) so it must be accepted that Winkler and Quimby are very nearly equivalent observers.

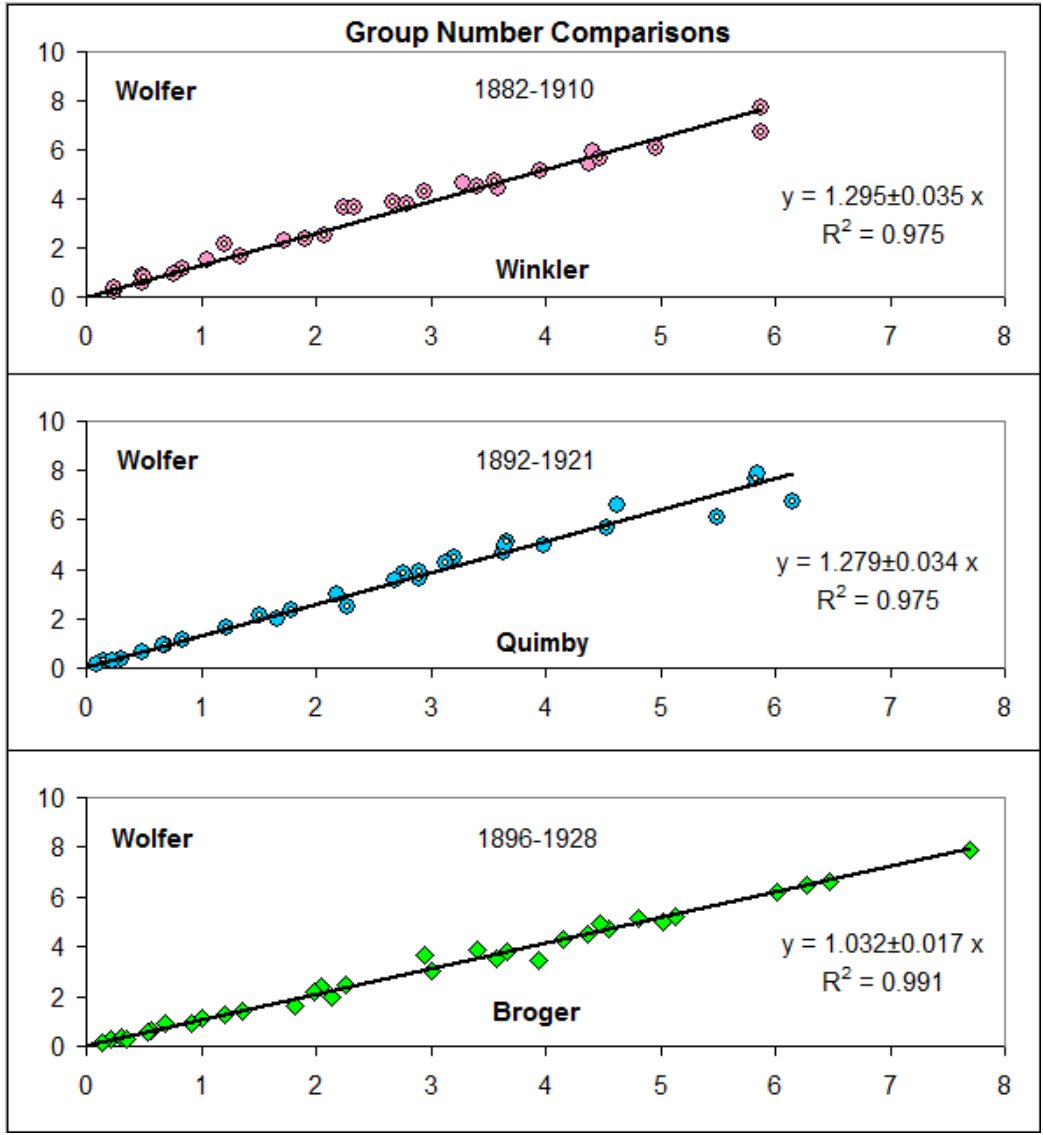

**Figure 1.** (Top) The average number of groups per day for each year 1882-1910 for observer Winkler compared to the number of groups reported by Wolfer. (Middle) The average number of groups per day for each year 1892-1921 for observer Quimby compared to the number of groups reported by Wolfer. Symbols with a small central dot mark common years between Winkler and Quimby. (Bottom) The average number of groups per day for each year 1896-1928 for the Zürich observer Broger compared to the number of groups reported by Wolfer. The slope of the regression line and the coefficient of determination $R^2$ are indicated on each panel. The offsets for zero groups are not statistically significant.

For days when two observers have both made an observation, we can construct a 2D-map of the frequency distribution of the simultaneous daily observations of the group counts *occurrence(groups(Observer1), groups(Observer2))*, i.e. showing on how many days Observer1 reports G1 groups while Observer2 reports G2 groups, varying G1 and G2 from 0 to a suitable maximum. Figure 2 shows such maps for Winkler and Quimby (Observers1) versus Wolfer (Observer2). It is clear that the maps are very similar and 'well-behaved', with narrow ridges stretching along the regression lines.

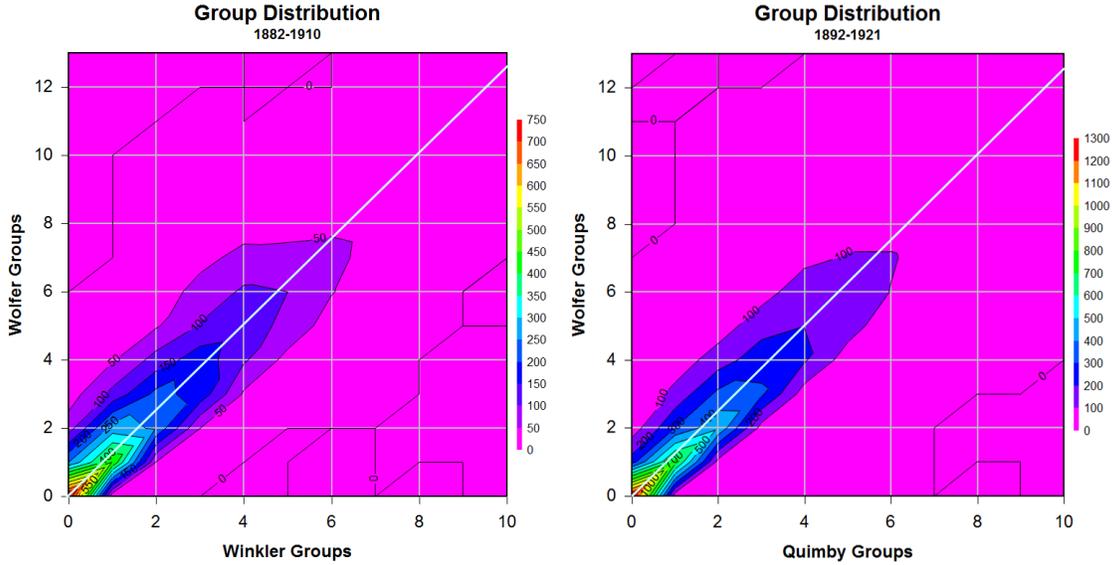

**Figure 2.** (Left) Distribution of simultaneous daily observations of group counts showing on how many days Winkler reported the groups on the abscissa while Wolfer reported the groups on the ordinate axis, e.g. when Winkler reported 5 groups, Wolfer reported 6 groups on 100 days during 1882-1910. (Right) Same, but for Quimby and Wolfer. The diagonal lines lie along corresponding group values determined by the daily *k'*-factors (≈1.25).

In Figure 3 (right panel) we plot the number of groups reported by Winkler against the number of groups reported by Quimby on the same day, to show that Winkler and Quimby are equivalent observers. The diagonal line marks equal frequency of groups reported by both observers.

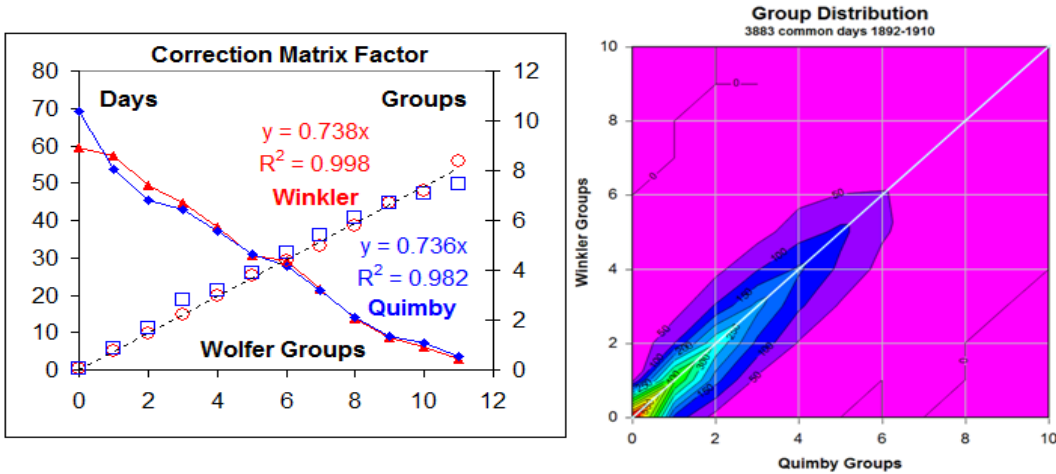

**Figure 3.** (Right) Distribution of simultaneous daily observations of group counts showing on how many days Quimby reported the groups on the abscissa while Winkler reported the groups on the ordinate axis, e.g. on days when Quimby reported 4 groups, Winkler also reported 4 groups on about 150 days during 1892-1910. (Left) The number of groups reported by Winkler (red circles) and by Quimby (blue squares) as a function of the number of groups reported by Wolfer on the same days. Also shown are the average number of days per year (left-hand scale) when those groups were observed (Winkler red triangles; Quimby blue diamonds). The factors are based on the 99% of the days where the group count is less than 12. Above that, the small-number noise is too large.

The 'Correction Factor' is the average factor to convert a daily group count by one observer to another. Figure 3 (left) shows that Winkler and Quimby have almost identical factors for conversion from Wolfer with almost identical distributions in time. This is again an indication that Winkler and Quimby are equivalent observers. If so, the yearly group numbers reported by the two observers should be nearly equal, which Figure 4 shows that they, as expected, are.

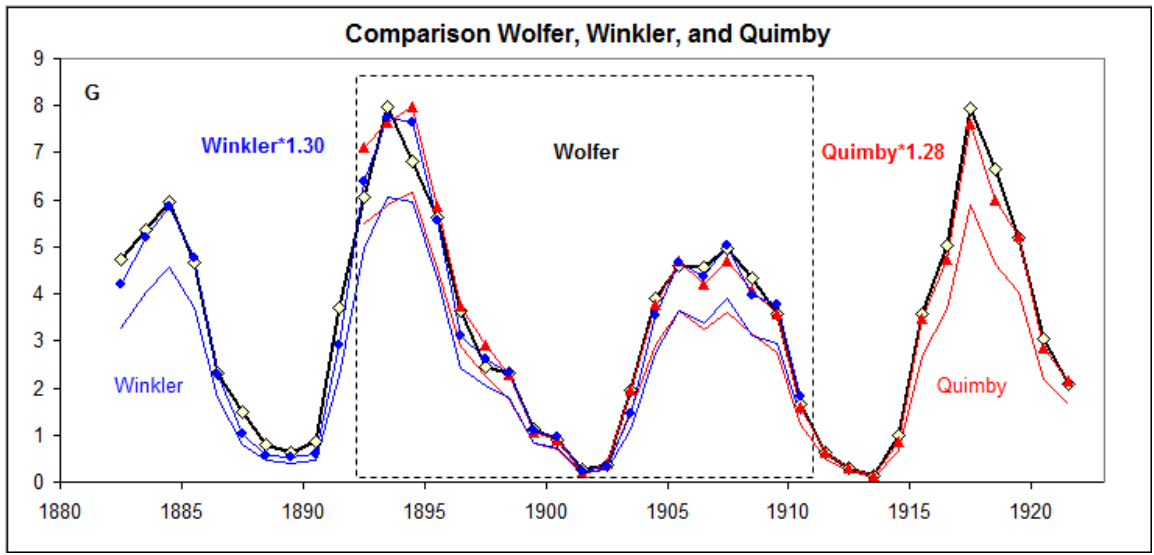

**Figure 4.** Yearly average reported group counts by Winkler (thin blue line) and Quimby (thin red line). The dashed line box outlines the years with common data. If we multiply the raw data by the $k'$-factors we get curves for Winkler (blue line with diamonds) and Quimby (red line with triangles) that should (and do) reasonably match the raw data for Wolfer (black line with light-yellow diamonds).

We have shown that Winkler and Quimby are equivalent observers and that their data multiplied by identical (within the errors) $k'$-factors reproduce the Wolfer observations.

**4. Broger and Wolfer are Equivalent Observers**

In Figure 1 we showed the average number of groups per day for each year 1896-1928 for the Zürich observer Broger compared to the number of groups reported by Wolfer. The $k'$-factor for Broger is unity within 2-$\sigma$, indicating that Broger and Wolfer are equivalent observers. For days when two observers have both made an observation, we can construct a 2D-map of the occurrence distribution of the 6778 simultaneous daily observations of counts during 1896-1928 similar to Figure 3. Figure 5 (right) shows the map for Broger versus Wolfer.

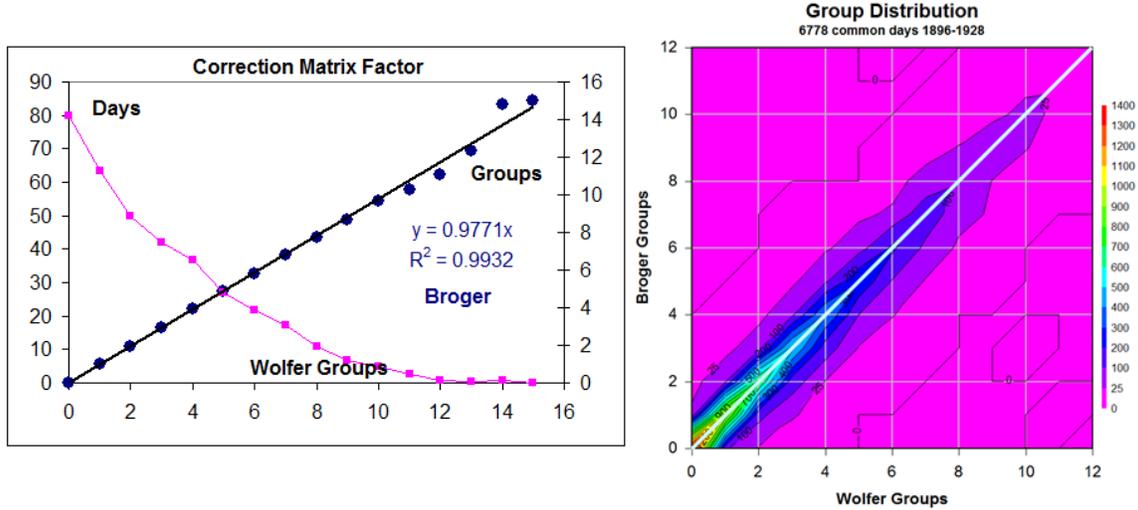

**Figure 5.** (Right) Distribution of simultaneous daily observations of group counts showing on how many days Wolfer reported the groups on the abscissa while Broger reported the groups on the ordinate axis, e.g. on days when Wolfer reported 4 groups, Broger also reported 4 groups on about 400 days during 1896-1928. (Left) The number of groups reported by Broger (dark-blue dots) as a function of the number of groups reported by Wolfer on the same days. Also shown are the average number of days per year (left-hand scale) when those groups were observed (pink squares).

Figure 5 shows that Broger and Wolfer have almost identical distributions in time. This is again an indication that Broger and Wolfer are equivalent observers. If so, the group numbers reported by the two observers should be nearly equal, which Figures 6 and 7 show that they, as expected, are.

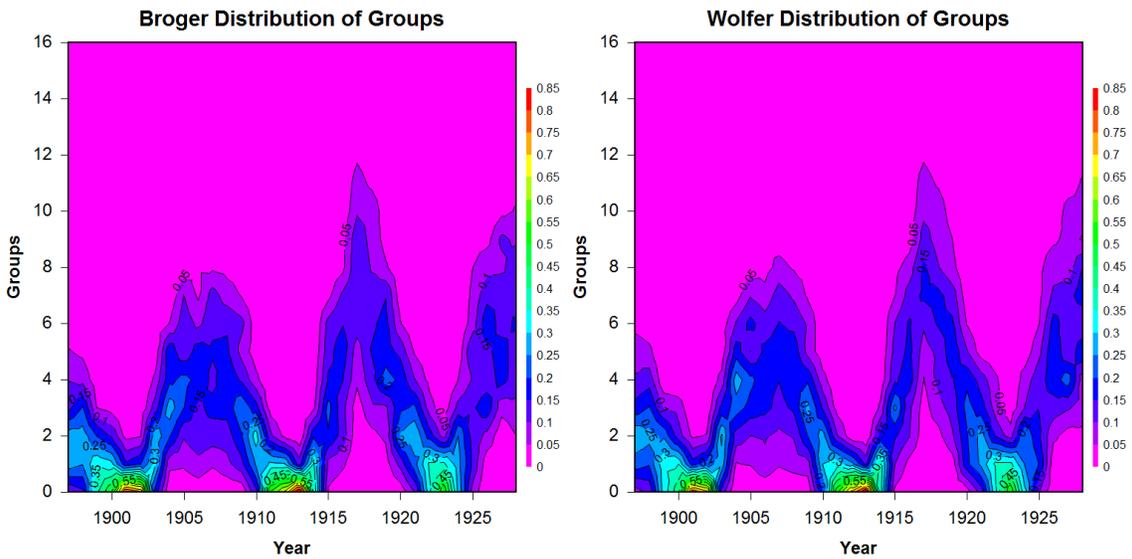

**Figure 6.** Distribution in time of daily observations of group counts showing the fraction of days per year Broger (left) and Wolfer (right) reported the groups on the ordinate axis).

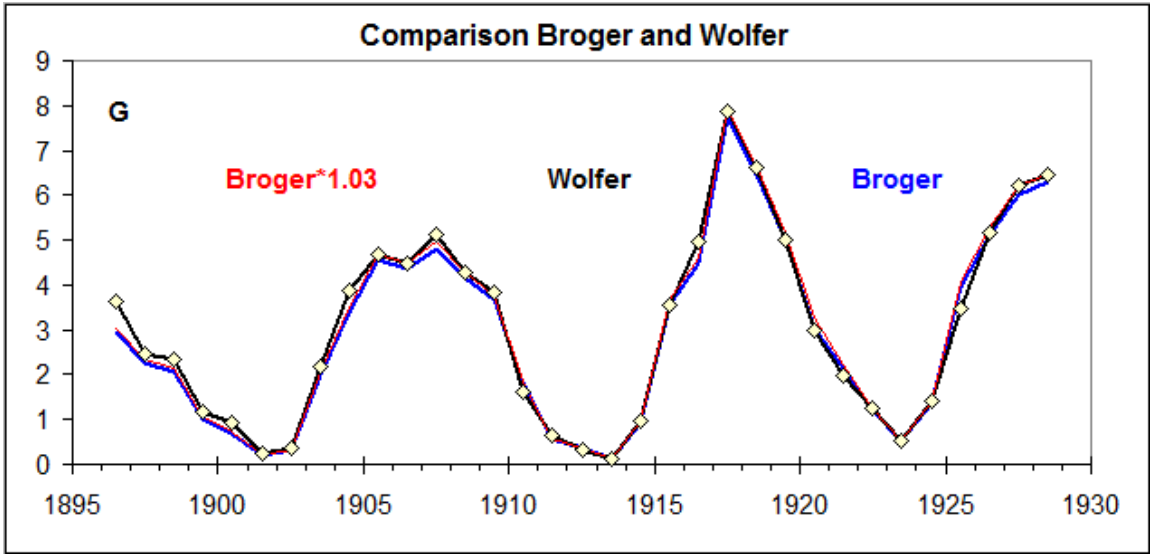

**Figure 7.** Yearly average reported group counts by Broger (blue line) and Wolfer (black line with light-yellow diamonds). If we multiply Broger's raw data by his *k'*-factor with respect to Wolfer we get the thin red line curve. There might be a hint of a slight learning curve for Broger for the earliest years.

We have shown that Broger and Wolfer are equivalent observers and that Broger's data reproduce the Wolfer observations. Combining the data in Figures 4 and 7 provides us with a firm and robust composite reconstruction of solar activity during the important transition from the 19$^{th}$ to the 20$^{th}$ centuries, Figure 8.

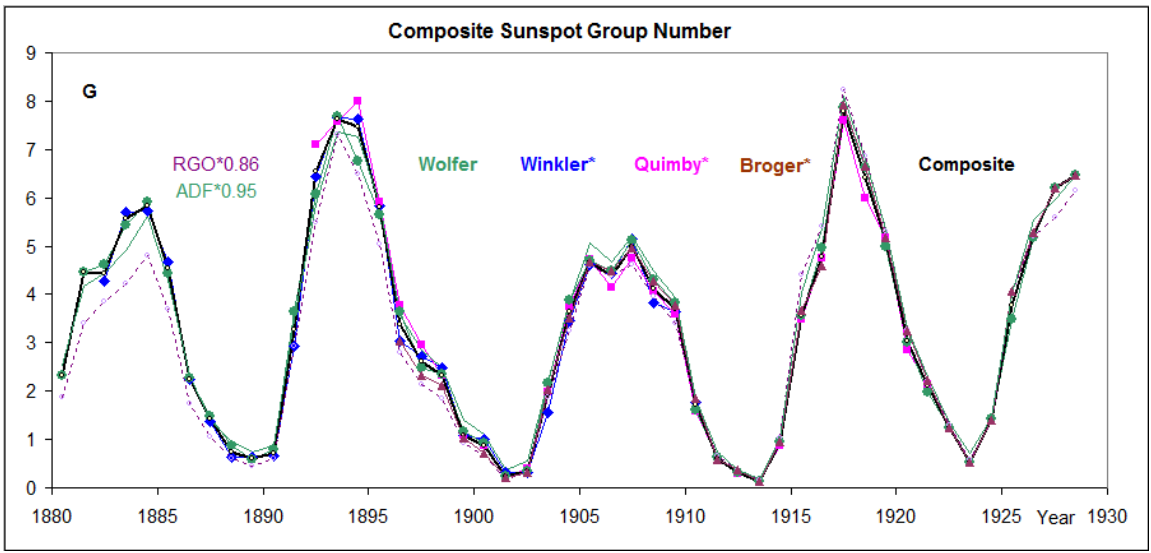

**Figure 8.** Composite Group Number series from Wolfer (green dots), Winkler (blue diamonds), Quimby (pink squares), and Broger (purple triangles). The dashed line shows the RGO (Royal Greenwich Observatory) group number scaled by a factor 0.86

derived from a fit with Wolfer spanning 1901-1928. The thin green line without symbols shows the ADF-based values from Willamo et al. [2017] scaled to fit Wolfer.

The consistency between Wolfer, Broger*, Quimby*, and Winkler*[5] throughout the years 1880-1928 suggests that there have been no systematic long-term drifts in the Composite. On the other hand, the well-known deficit for RGO before about 1890 is clearly evident. The ADF-based values seem at first blush to match the Composite reasonably well. Unfortunately, the agreement is spurious as we shall show in the following sections.

## 5. The ADF Observational Threshold

The ADF-method [Willamo et al., 2017] is based on the assumption that the 'quality' of each observer is characterized by his/her acuity given by an observational threshold area $S^6$, on the solar disk of all the spots in a group. The threshold (all sunspot groups with an area smaller than that were considered as not observed) defines a calibration curve derived from the cumulative distribution function (CDF) of the occurrence in the reference dataset (RGO) of months with the given ADF. A family of such curves is produced for different values of $S$. The observational threshold for each observer is defined by fitting the actual CDF curve of the observer to that family of calibration curves. The best-fit value of $S$ and its 68% ($\pm 1\sigma$) confidence interval were defined by the $\chi^2$ method with its minimum value corresponding to the best-fit estimate of the observational threshold. Table 1 gives the thresholds for the observers considered in this article.

**Table 1.** The columns are: the name of the observer, the Fraction of Active Days, the lower limit of $S$ for the 68% confidence interval, the observational threshold area $S$ in millionth of the solar disk, the upper limit of $S$, and the observer's code number in the Vaquero et al. [2016] database. (From Willamo et al., [2017]).

| Observer | ADF % | S low | S μsd | S high | Code |
|---|---|---|---|---|---|
| RGO | 86 | - | **0** | - | 332 |
| Spörer | 86 | 0 | **0** | 2 | 318 |
| Wolfer | 77 | 1 | **6** | 11 | 338 |
| Broger | 78 | 5 | **8** | 11 | 370 |
| Weber | 81 | 20 | **25** | 31 | 311 |
| Shea | 80 | 20 | **25** | 31 | 295 |
| Quimby | 73 | 17 | **23** | 31 | 352 |
| Winkler | 75 | 51 | **60** | 71 | 341 |

## 6. Does the ADF-method Work for Equivalent Observers?

We have shown above (Section 3 and 4) that pairs of Equivalent Observers (same observational thresholds or same k'-factors) saw and reported the same number of groups (Figures 4 and 7). As a minimum one must demand that the group numbers determined using the ADF-method also match the factually observed equality of a pair of equivalent observers. If the ADF-method yields significant difference between what two equivalent observers actually reported, we cannot expect the method to give correctly calibrated results for those two observers. We assert that this is true regardless of the inner workings and irreproducible computational details of the ADF-method (or any method for that matter).

---
[5] The asterisks denote the raw values multiplied by the *k'*-factor.
[6] Simplified form of the $S_S$ used by Willamo et al. [2017].

### 6.1. ADF Fails for Quimby and Winkler

Figure 9 shows the ADF-based group numbers (from Willamo et al. [2017]) for the Equivalent Observers Quimby and Winkler.

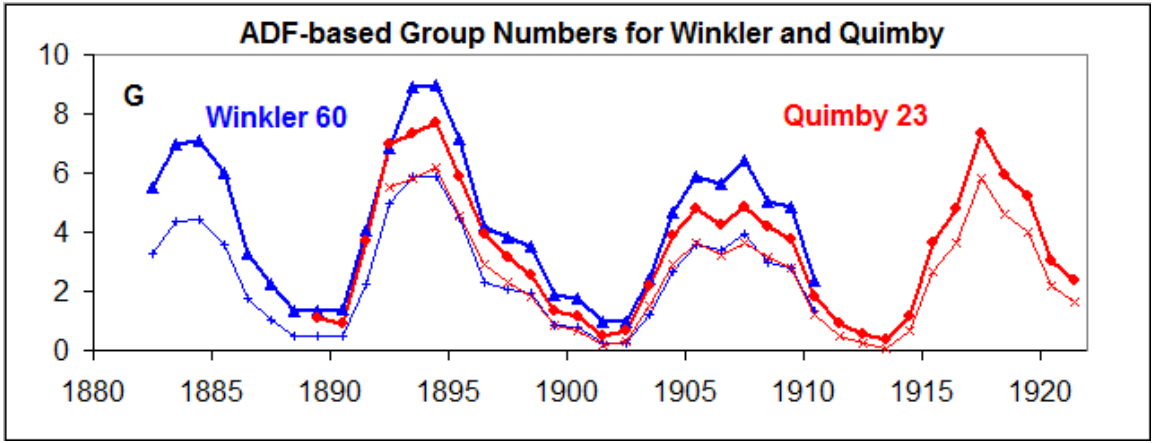

**Figure 9.** ADF-based group numbers for Winkler ($S = 60$, blue triangles) and Quimby ($S = 23$, red dots). The raw, actually observed group numbers for Winkler ($k' = 1.3$, blue plusses) and Quimby ($k' = 1.3$, red crosses) are shown below the ADF-based curves.

It should be evident that ADF-method fails to produce the expected nearly identical counts observed by these two equivalent observers, not to speak about the large discrepancy (60 vs. 23) in the $S$ threshold areas.

### 6.2. ADF Fails for Broger and Wolfer

Figure 10 shows the ADF-based group numbers (from Willamo et al. [2017]) for the Equivalent Observers Broger and Wolfer.

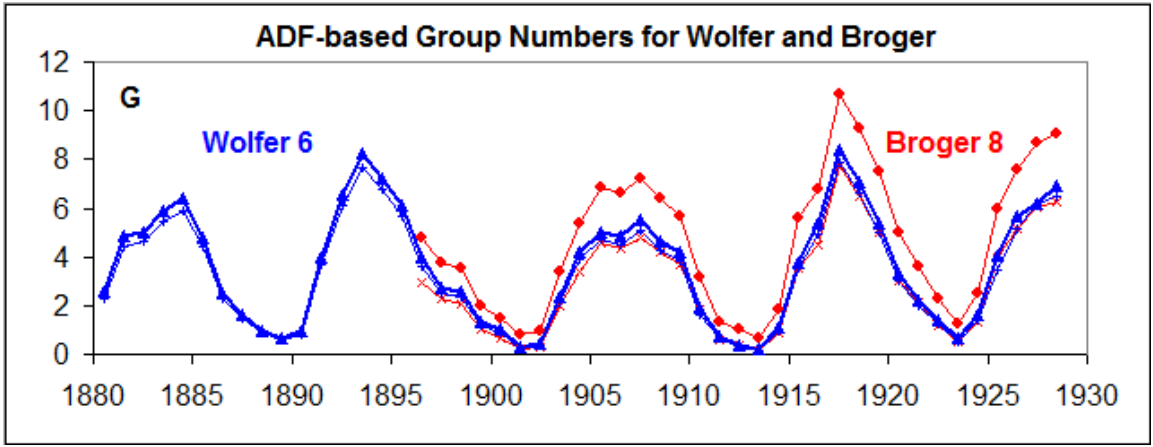

**Figure 10.** ADF-based group numbers for Wolfer ($S = 6$, blue triangles) and Broger ($S = 8$, red dots). The raw, actually observed group numbers for Wolfer ($k' = 1.0$, blue plusses) and Broger ($k' = 1.0$, red crosses) are shown below the ADF-based curves.

It should be evident that ADF-method fails to produce the expected nearly identical counts observed by these two equivalent observers, in spite of the nearly identical $S$ threshold areas.

### 6.3. ADF Fails for Weber and Shea

Table 1 shows that Heinrich Weber (observed 1859-1883) and Charles Shea (observed 1847-1866, 5538 drawings reduced by Hoyt & Schatten) should also be equivalent observers because they have identical $S$ values of 25. Figure 11 shows the ADF-based group numbers (from Willamo et al. [2017]) and the actual observed group numbers for Weber and Shea.

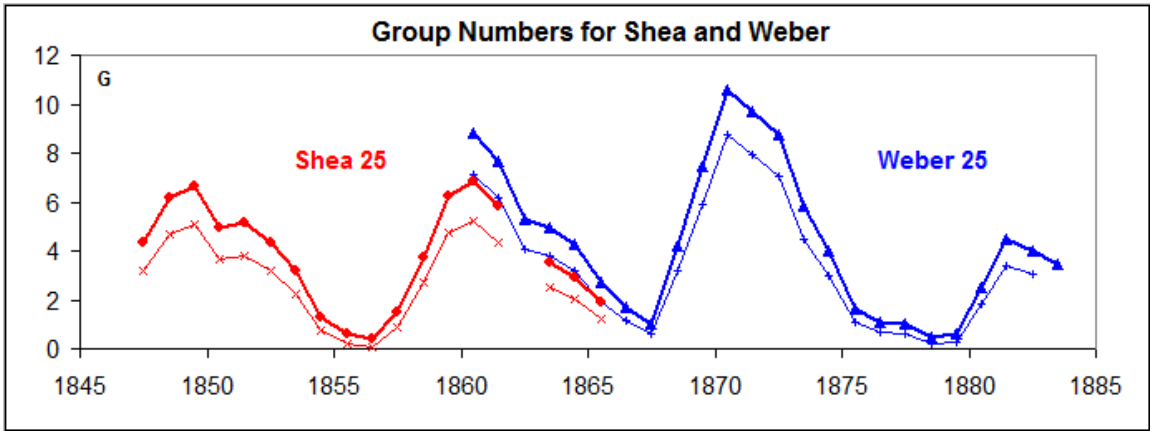

**Figure 11.** ADF-based group numbers for Weber ($S = 25$, blue triangles) and Shea ($S = 25$, red dots). The raw, actually observed group numbers for Weber (blue plusses) and Shea (red crosses) are shown below the ADF-based curves.

It should be evident that the ADF-method fails to produce the expected nearly identical counts observed by these two observers with identical $S$ threshold areas. In addition, the actual observations are not consistent with equal $S$ values since Weber reported 40% more groups than Shea. Data for 1862 are missing from the database. The observations by Shea are preserved in the Library of the Royal Astronomical Society (London) and bear re-examination.

### 6.4. ADF Fails for Spörer and RGO

Table 1 shows that Gustav Spörer (1822-1895, observed 1861-1893) and the Greenwich observers (1884-1976) are both 'perfect observers' [Willamo et al., 2017] since their $S$ value is zero[7]. We should therefore expect that they should observe and report nearly identical yearly values of the sunspot group numbers, as they have the same observational threshold and no groups should be missed. Figure 12 shows that perfect observer Spörer does not at all match the other perfect observer RGO.

---

[7] The data for 1879 for Spörer are anomalously high because all days with zero groups were entered as missing in the Hoyt & Schatten catalog. This may have influenced slightly the determination of $S$.

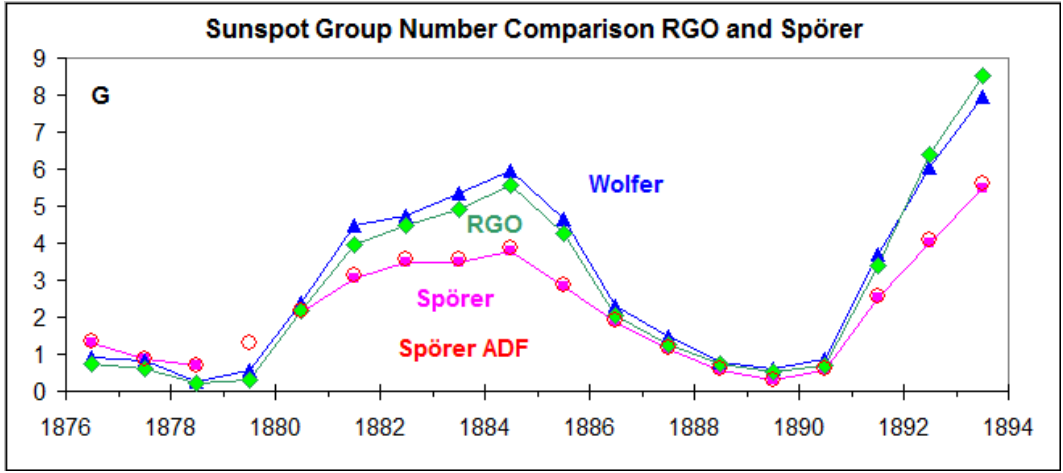

**Figure 12.** Annual values of the observed Sunspot Group Numbers for Spörer (pink squares), RGO (green diamonds), Wolfer (blue triangles), and for Spörer computed by Willamo et al. [2017] using the ADF-method (red open circles). Note that Wolfer ($S = 6$) and RGO ($S = 0$) are pretty close, as expected, showing that for these years the drift of RGO was not so significant, yet.

Spörer needs to be scaled up by a factor 1.45 to match RGO, so can hardly be deemed to be a 'perfect observer' as determined by the ADF-method. More details on this issue can be found in Svalgaard [2017].

## 7. The Problem with Zero Groups

Even if we compare two equivalent observers there will be a spread in the values. If one observer sees, say, four groups on a given day, the other observer will often observe a different number, because of variable seeing and of small groups emerging, merging, splitting, or disappearing at different times for the two observers. So there is a 'point-spread function' with a round hill of width typically one to two groups, centered on the chosen group number value, Figure 13.

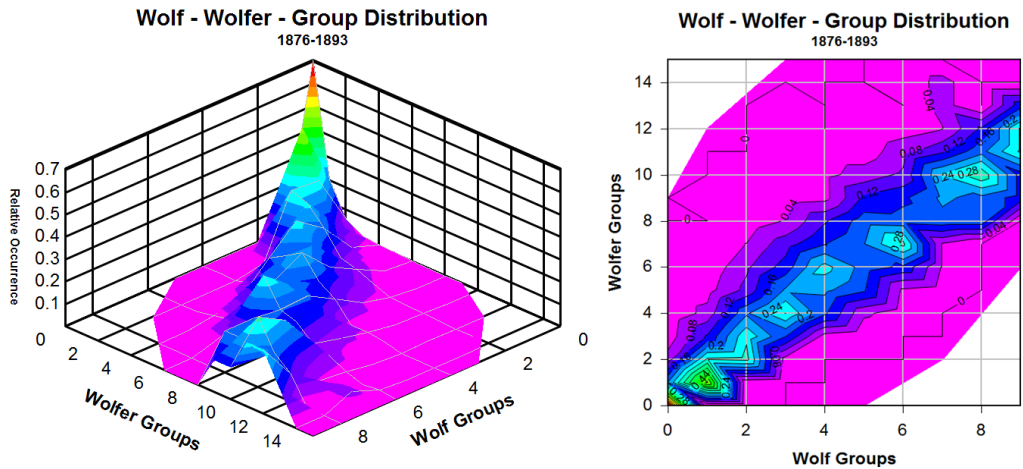

**Figure 13.** The distribution of daily values of the observed Sunspot Group Numbers for Wolfer for each bin of Wolf's group number, normalized to the sum of all groups in that bin. (Left) A 3D view of the 'hills' for each bin. (Right) A contour plot of the distribution.

So, in general, there will be a neighborhood in the distribution around a given group number 'hill' where some group numbers are a bit larger and some are a bit smaller than the top-of-the-hill number. This holds for all bins *except* for the zero bin, because there are no negative group numbers. As a result, the other observer's average group number for the first observer's zero bin will be artificially too high. This fundamental flaw can be seen in the ADF-series for all observers, rendering the ADF-values generally too high for low activity. The purpose of the ADF-method is to bring all observers considered onto the same scale. As Figure 14 shows this goal is not realized for low solar activity.

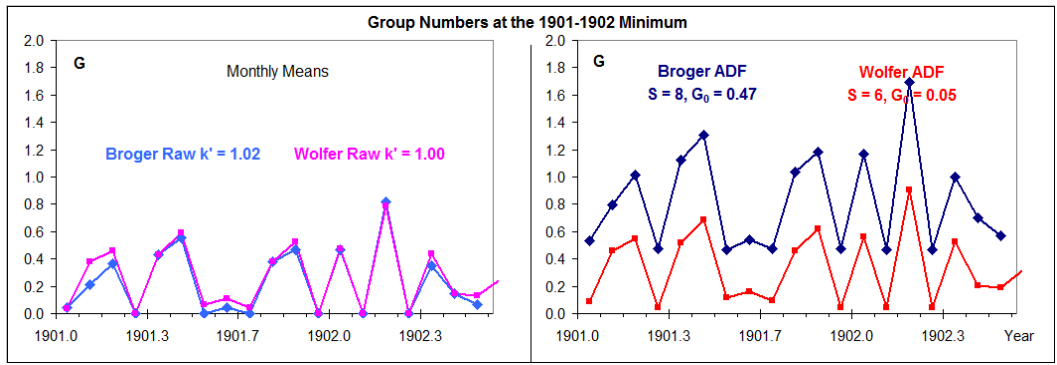

**Figure 14.** (Left) The monthly mean Group Numbers observed by the equivalent observers Broger (light-blue diamonds) and Wolfer (pink squares) during the deepr solar minimum 1901.0-1902.6. (Right) The Group Numbers for Broger (dark-blue diamonds) and Wolfer (red squares) computed by Willamo et al. [2017] using the ADF-method. The artificial offset for Broger (0.47) is particularly egregious for $G_{\text{Wolfer}} = 0$.

## 8. Conclusion

We have identified several pairs of 'equivalent' observers and shown that the group numbers computed using the ADF-method do not reproduce the equality of the group numbers expected for equivalent observers, rendering the vaunted[8] ADF-methodology suspect and not reliable nor useful for studying the long-term variation of solar activity. We suggest that the claim [Willamo et al., 2017] that their "new series of the sunspot group numbers with monthly and annual resolution, […] is forming a basis for new studies of the solar variability and solar dynamo for the last 250 years" is premature, and, if their series is used, will hinder such research. It is incumbent on the community to resolve this issue [Cliver, 2016] so progress can be made, not just in solar physics, but in the several diverse fields using solar activity as input.

---

[8] frequentative of Latin vanare: "to utter empty words"

## Acknowledgements

We thank Ilya Usoskin for the data files communicated to Laure Lefèvre; L.S. thanks Stanford University for continuing support.

## Disclosure of Potential Conflicts of Interest

The authors declare that they have no conflicts of interest.